\documentclass[a4paper,11pt]{article}
\usepackage{pos}
\usepackage{orcidlink}

\title{Taste-splittings of staggered, Karsten-Wilczek and Borici-Creutz fermions under gradient flow in 2D}
\ShortTitle{Taste-splittings of staggered, KW and BC fermions under gradient flow in 2D}

\author[a]{Stefano Capitani}
\author*[a,b]{Stephan D\"urr\,\orcidlink{0000-0001-5168-5669}\,}

\affiliation[a]{Physics Department, University of Wuppertal, 42119 Wuppertal, Germany}
\affiliation[b]{IAS/JSC, Forschungszentrum J\"ulich, 52425 J\"ulich, Germany}

\emailAdd{capitani\,(AT)\,uni-wuppertal$\mbox{.}$de}
\emailAdd{   duerr\,(AT)\,uni-wuppertal$\mbox{.}$de}

\abstract{Karsten-Wilczek and Borici-Creutz fermions show a near-degeneracy
of the $2$ species involved, similar to the $2^{d/2}$ species of staggered fermions.
Hence in $d=2$ dimensions all three formulations happen to be minimally doubled (two species).
This near-degeneracy shows up both in the eigenvalue spectrum of the respective Dirac operator and
in spectroscopic quantities (e.g.\ the pion mass), but in the former case it is easier to quantify.
We use the quenched Schwinger model to determine the low-lying eigenvalues of these fermion operators at a fixed gradient flow
time $\tau$ (either in lattice units or in physical units, hence keeping either $\tau/a^2$ or $e^2 \tau$ fixed at all $\beta$).}

\FullConference{%
The 41st International Symposium on Lattice Field Theory (LATTICE2024)\\
28 July - 3 August 2024\\
Liverpool, UK}

\newcommand{\dal}{{\sqcap\!\!\!\!\sqcup}}

\newcommand{\be}{\beta}

\newcommand{\de}{\delta}
\newcommand{\ep}{\epsilon}

\newcommand{\la}{\lambda}

\newcommand{\ta}{\tau}

\newcommand{\vp}{\varphi}

\newcommand{\bdm}{\begin{displaymath}}
\newcommand{\edm}{\end{displaymath}}
\newcommand{\bea}{\begin{eqnarray}}
\newcommand{\eea}{\end{eqnarray}}
\newcommand{\beq}{\begin{equation}}
\newcommand{\eeq}{\end{equation}}

\newcommand{\mr}{\mathrm}
\newcommand{\ri}{\mr{i}}
\newcommand{\Nf}{N_{\!f}}

\long\def\begincomment#1\endcomment{}

\begin{document}

\maketitle


\section{Introduction}

Taste splittings are an unwanted effect -- a lattice artefact or ``cut-off effect''.
They are genuine to any lattice fermion action involving more than one species (a.k.a.\ ``doubled action'').

\begin{figure}[!b]
\includegraphics[width=0.5\textwidth]{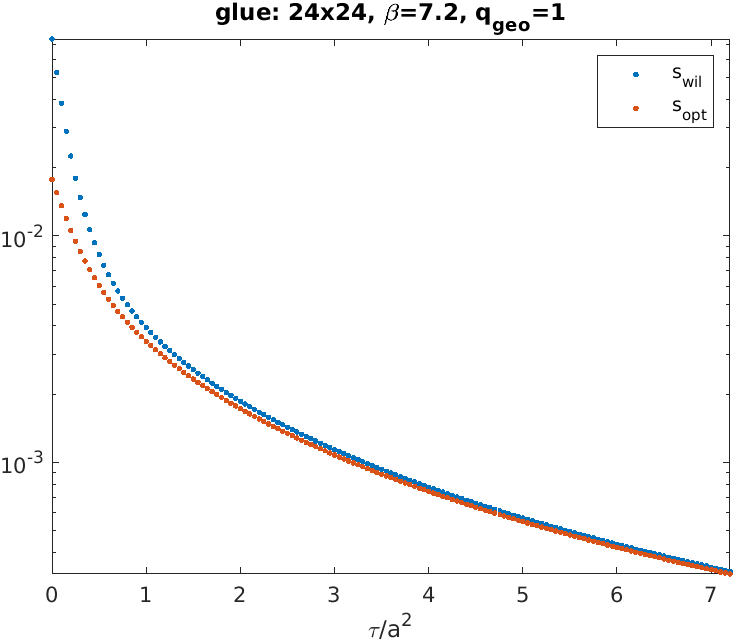}%
\includegraphics[width=0.5\textwidth]{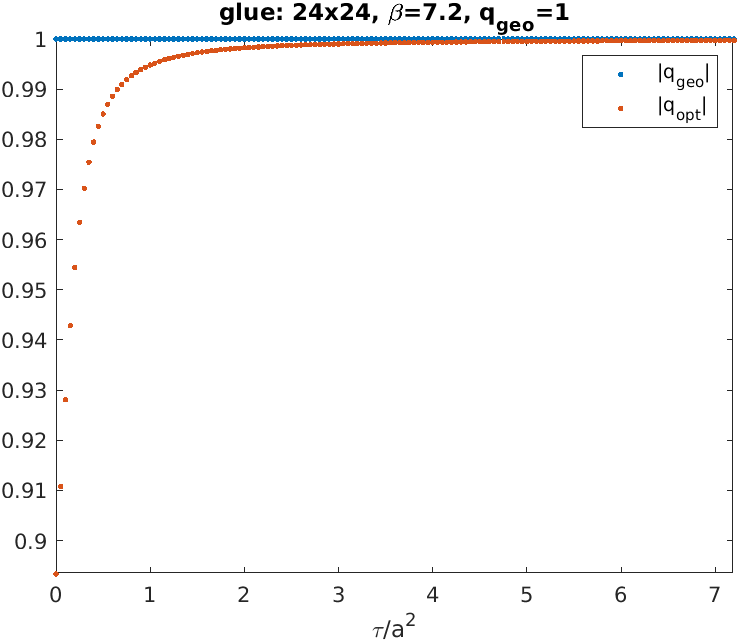}%
\vspace*{-4pt}
\caption{\label{fig:glue1}
Effect of the gradient flow on two gluonic actions (left) and on two topological charges (right).}
\end{figure}

For instance staggered fermions involve $2^{d/2}$ species in $d$ space-time dimensions (i.e.\ 2 species or ``tastes'' in 2D, and 4 in 4D).
This is visible in the eigenvalue spectrum of $aD_\mr{stag}$; instead of one continuum eigenvalue one finds a \emph{pair} (in 2D) or
a \emph{quartet} (in 4D) of near-degenerate eigenvalues on a representative gauge background $U_\mu(n)$ \cite{Follana:2004sz,Durr:2004as}.
Accordingly, in a dynamical simulation with one field of $aD_\mr{stag}$ the well-known rooting procedure effectively replaces
each pair (quartet) by the geometric mean of the two (four) multiplet eigenvalues in 2D (4D).

More recently, Karsten-Wilczek (KW) \cite{Karsten:1981gd,Wilczek:1987kw} and Borici-Creutz (BC) \cite{Creutz:2007af,Borici:2007kz} fermions were proposed, since they entail only two species.
This is just the minimum number required by the Nielsen-Ninomyia theorem (and hence the same number in 2D and 4D).
As a result, in 4D one can simulate $\Nf=2$ QCD with KW or BC fermions without using the rooting trick.

Specifically in 2D, not just KW and BC fermions, but also staggered fermions happen to be ``minimally doubled''.
Accordingly, the Schwinger model (QED in 2D) is well suited to compare the taste-breaking effects of these three fermion formulations to each other.
There are two options for addressing the taste breaking effects.
One may determine spectrosopic quantities like $a^2M_{\pi,V}^2-a^2M_{\pi,P}^2$, where the second subscript indicates the taste structure of the pion.
Or one may measure the above mentioned eigenspectra and determine the splitting within each pair.

Today it is common practice to evaluate the Dirac operator $aD$ on a gauge background $V$ which is derived from the actual
configuration $U$ via a few steps of stout smearing \cite{Morningstar:2003gk} or some gradient flow evolution \cite{Luscher:2010iy,Luscher:2011bx}.
At first sight the difference between these smoothings procedures is small, since the correspondence
$n_\mr{stout}\rho_\mr{stout}=\tau_\mr{flow}/a^2$ (see e.g.\ \cite{Ammer:2024hqr} and references therein) says that
the flow time in lattice units (r.h.s.) equals the cumulative sum of the stout parameters used (l.h.s.).

The effect of 1 or 3 stout steps on the eigenvalues of $aD_\mr{stag}$, $aD_\mr{KW}$ and $aD_\mr{BC}$ has been investigated in Ref.\,\cite{Ammer:2024yro}.
Here we take first steps towards exploring the effect of the gradient flow.


\section{Effect of the gradient flow on gluonic quantities}

\begin{figure}[!t]
\includegraphics[height=0.43\textwidth]{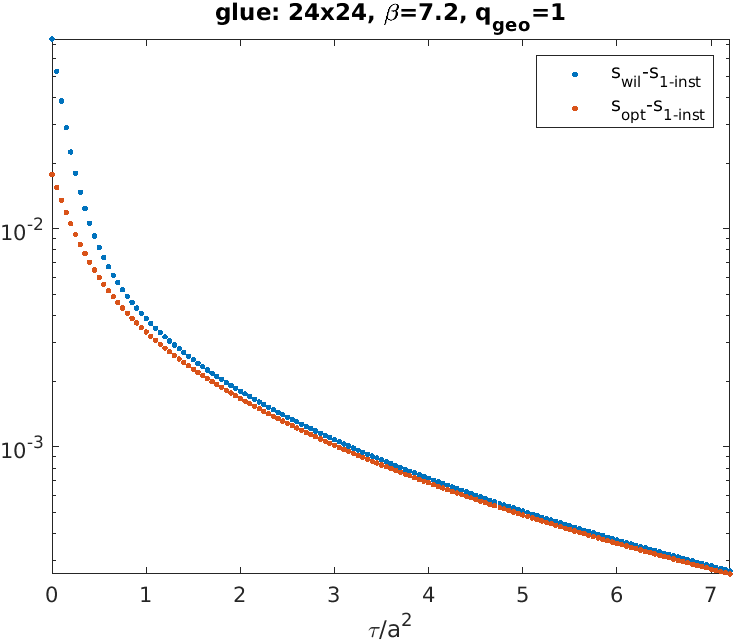}%
\includegraphics[height=0.43\textwidth]{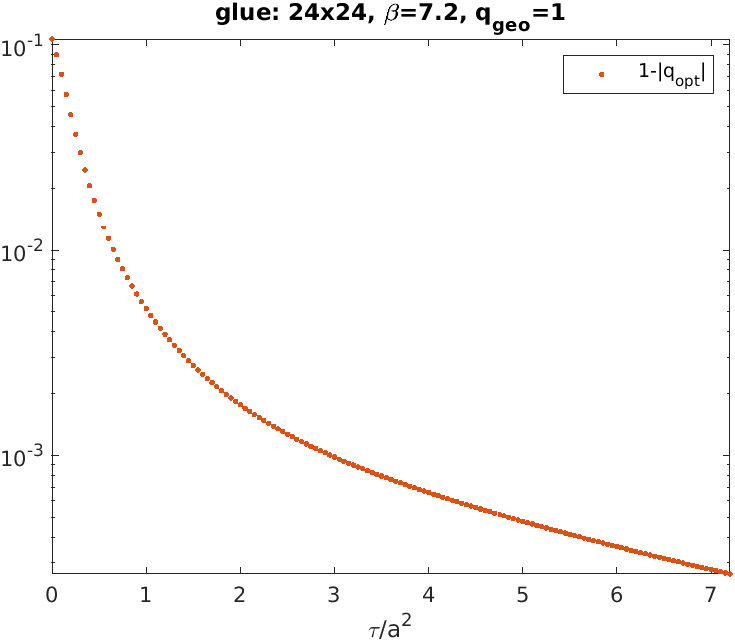}%
\vspace*{-4pt}
\caption{\label{fig:glue2}
Same data as in Fig.\,\ref{fig:glue1}, after subtracting the respective (analytically known) 1-instanton values.}
\end{figure}

In the two-dimensional $U(1)$ theory one defines $U_\dal(n)=U_1(n) U_2(n+\hat{1}) U_1^*(n+\hat{2}) U_2^*(n)$
where $n=(x,y)$ is the lattice site and $U_\mu^*(n)$ the complex conjugate of $U_\mu(n)$.
Using the parametrization $U_\mu(n)=e^{\ri\vp_\mu(n)}$ one may write $U_\dal(n)=e^{\ri\vp_\dal(n)}$ with
the plaquette angle, mapped to the interval $[-\pi,\pi[$, given by $\vp_\dal(n)\equiv\mr{mod}(\vp_1(n)+\vp_2(n+\hat{1})-\vp_1(n+\hat{2})-\vp_2(n)+\pi,2\pi)-\pi$.

The Wilson action is $S_\mr{wil}[U]=\be\sum_n\{1-\mr{Re}\,U_\dal(n)\}=\be\sum_n\{1-\cos\vp_\dal(n)\}$ and
another option is $S_\mr{opt}[U]=\be\sum_n\frac{1}{32}[\sin\vp_\dal(n)+\sin\vp_\dal(n-\hat{1})+\sin\vp_\dal(n-\hat{1}-\hat{2})+\sin\vp_\dal(n-\hat{2})]^2$.
Two definitions of the topological charge are in common use,
the geometric charge $q_\mr{geo}[U]=\frac{1}{2\pi}\sum_n\mr{Im}\log U_\dal(n)=\frac{1}{2\pi}\sum_n\vp_\dal(n)\in\mathbb{Z}$
and optionally $q_\mr{opt}[U]=\frac{1}{2\pi}\sum_n\mr{Im}\,U_\dal(n)=\frac{1}{2\pi}\sum_n\sin\vp_\dal(n)\in\mathbb{R}$.

We choose a thermalized gauge configuration at $(\be,L/a)=(7.2,24)$, and plot $s(\ta)/\be$ and $q(\ta)$ versus the flow time $\ta/a^2$ in the interval $[0,7.2]$ in Fig.\,\ref{fig:glue1}.
It seems that $s_\mr{wil}-s_\mr{opt}\to0$ and $q_\mr{geo}-q_\mr{opt}\to0$ for large $\ta/a^2$, as expected.
Fortunately, the $q$-instanton configuration in the Schwinger model is known analytically \cite{Smit:1987fq};
its action is $s_{q-\mr{inst}}/\be=1-\cos(\frac{2\pi q}{N_x N_y})$.
Hence, by subtracting from either observable its $\ta/a^2=\infty$ value, we can study the asymptotic ascent.
Fig.\,\ref{fig:glue2} suggests that for $s$ and $q$ the asymptotic value is assumed \emph{exponentially} in the flow time $\ta/a^2$.

\begin{table}[!b]
\centering
\begin{tabular}{|c|ccccccc|}
\hline
$\be$              & 3.2 & 5.0 & 7.2 & 12.8 & 20.0 & 28.8 & 51.2 \\
$L/a$              & 16  & 20  & 24  & 32   & 40   & 48   & 64   \\
$\ta_\mr{max}/a^2$ & 3.2 & 5.0 & 7.2 & 12.8 & 20.0 & 28.8 & 51.2 \\
\hline
\end{tabular}
\vspace*{-2pt}
\caption{\label{tab:planning}
Parameters for matched lattices and gradient flow times. The volume in physical units $(eL)^2=(L/a)^2/\be$ is
always $80$, the maximum flow time in physical units $e^2\ta_\mr{max}=\ta_\mr{max}/(a^2\be)$ is always $1$.}
\end{table}

We checked the effect that larger/smaller boxes at the same coupling $\be$ have; we found no significant change.
In the Schwinger model varying the lattice spacing $a$ at fixed physical box size $L$ is simple,
if $a$ is set through the dimensionful coupling $e$, since $\be=1/(ae)^2$.
This allows us to compile a list of matched lattices/flow-times before running any simulation, see Tab.\,\ref{tab:planning}.


\begin{figure}[!t]
\includegraphics[width=0.5\textwidth]{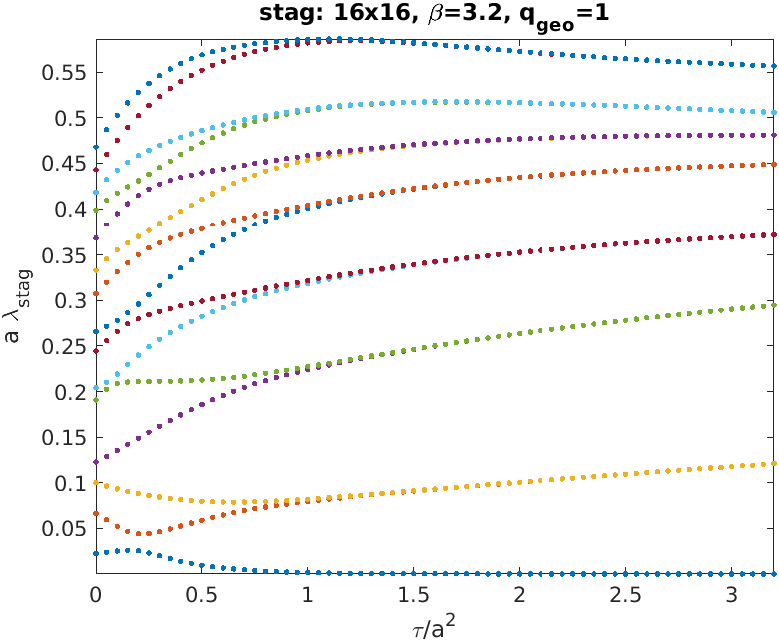}%
\includegraphics[width=0.5\textwidth]{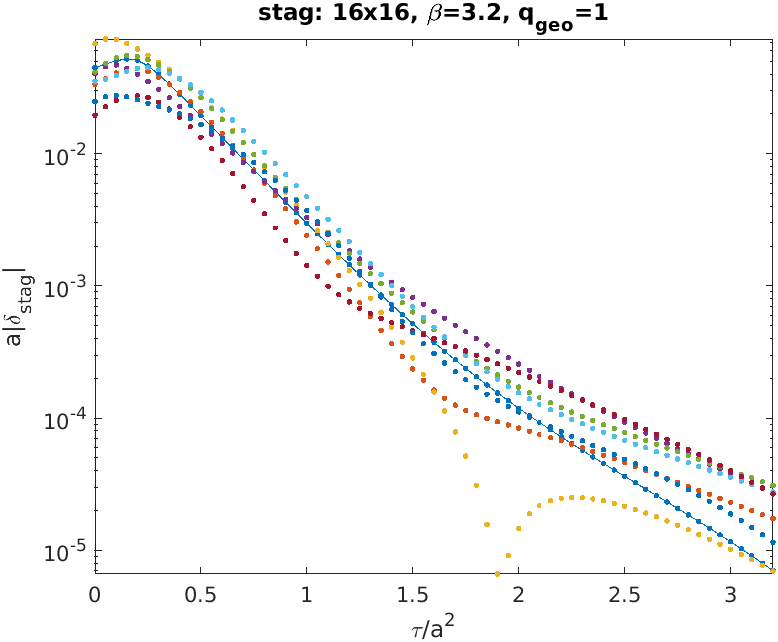}\\
\includegraphics[width=0.5\textwidth]{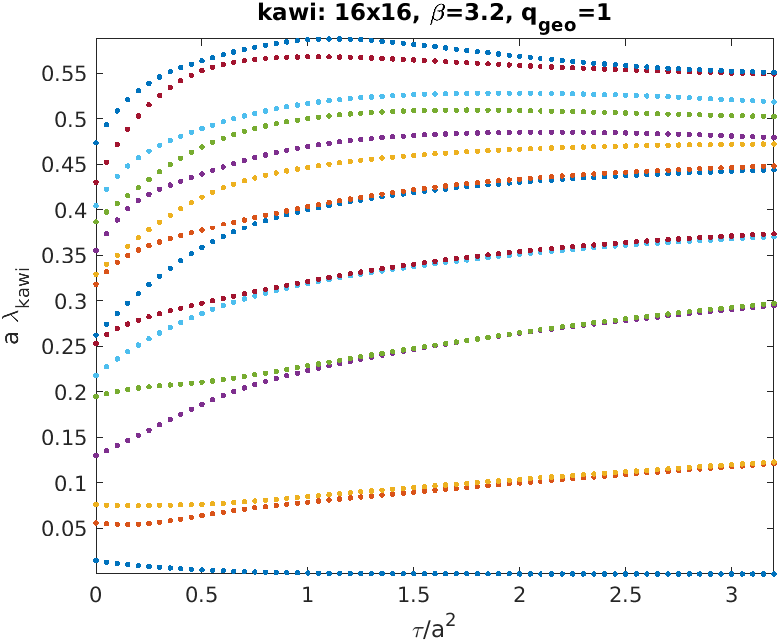}%
\includegraphics[width=0.5\textwidth]{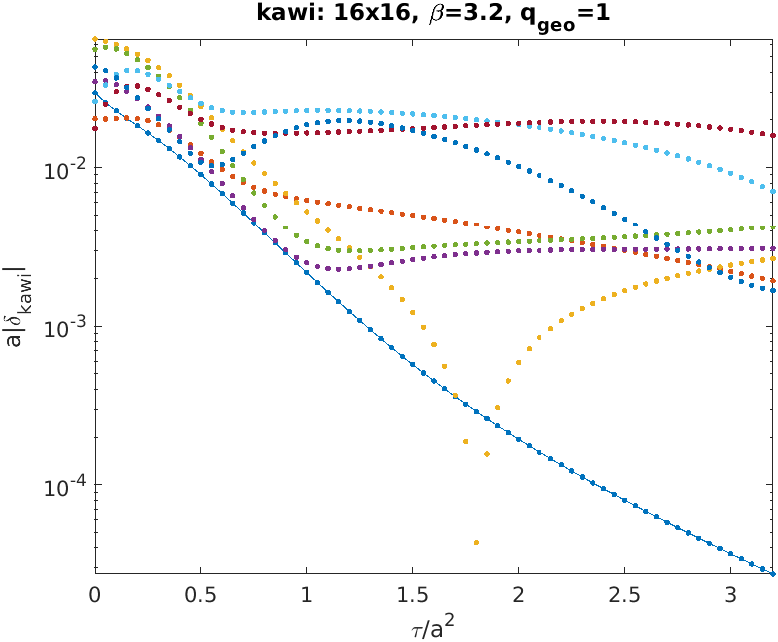}\\
\includegraphics[width=0.5\textwidth]{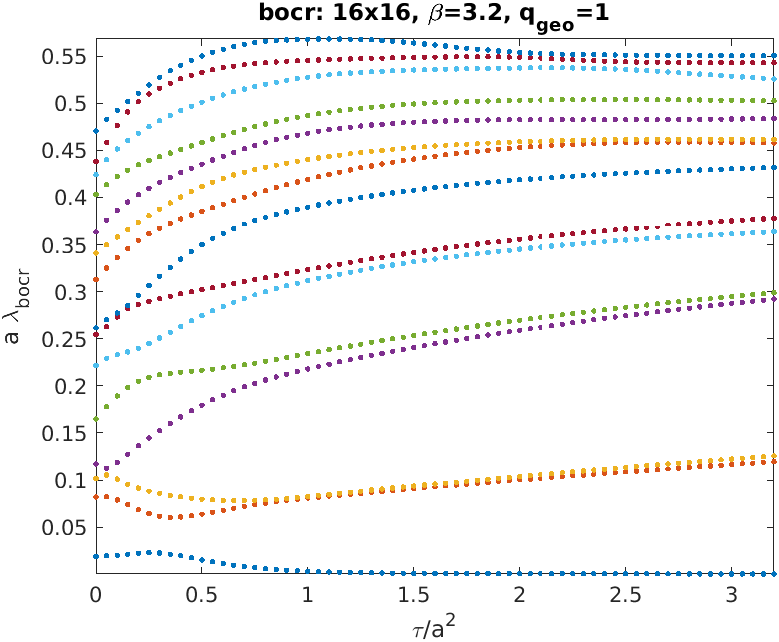}%
\includegraphics[width=0.5\textwidth]{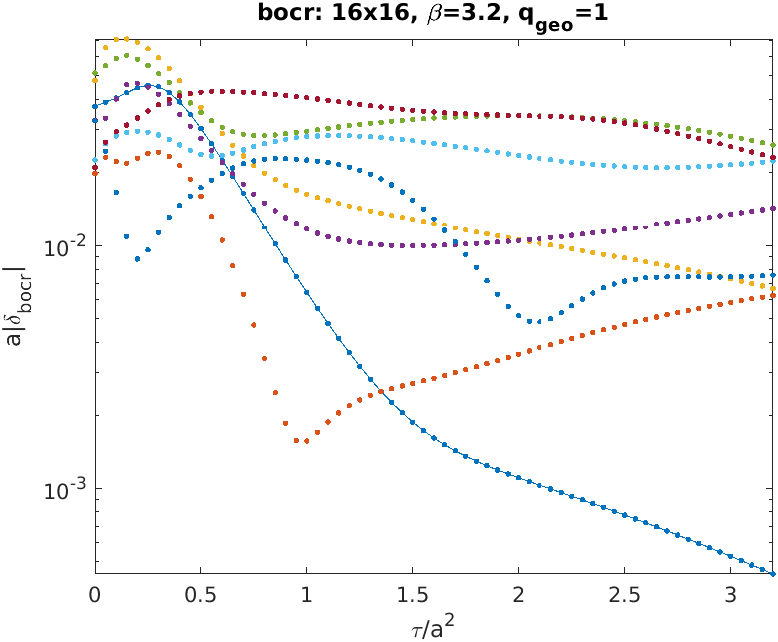}%
\vspace*{-4pt}
\caption{\label{fig:03p2}
Left: Upper half of the eigenvalues $\ri\la$ of $aD_\mr{stag}$ (top), $aD_\mr{KW}$ (middle) and $aD_\mr{BC}$ (bottom)
on a $\be=3.2$ configuration with $|q|=1$ versus the flow time $\ta/a^2$. Right: Taste splittings derived from these data.}
\end{figure}

\begin{figure}[!t]
\includegraphics[width=0.5\textwidth]{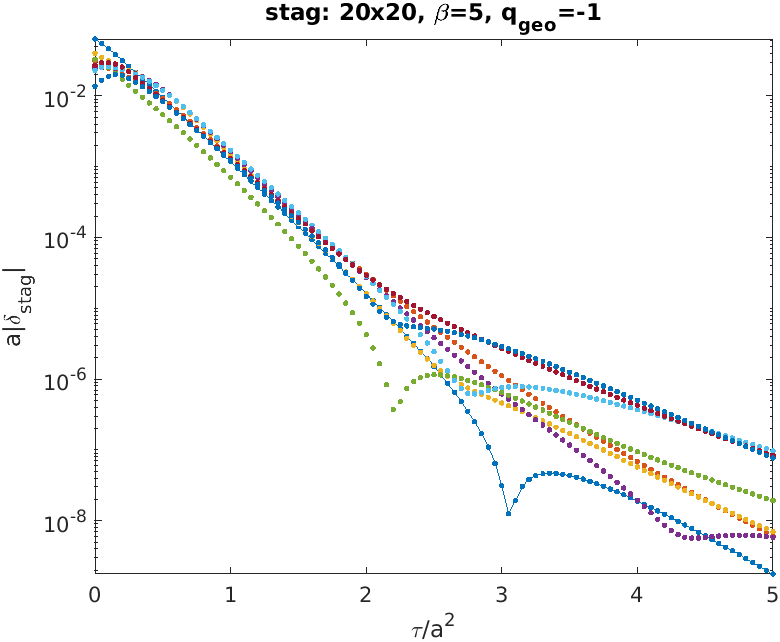}%
\includegraphics[width=0.5\textwidth]{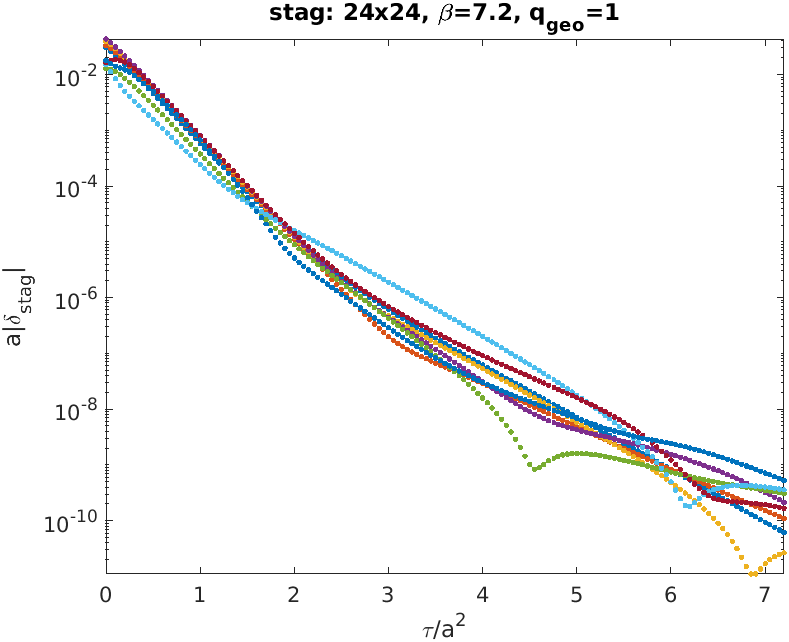}\\
\includegraphics[width=0.5\textwidth]{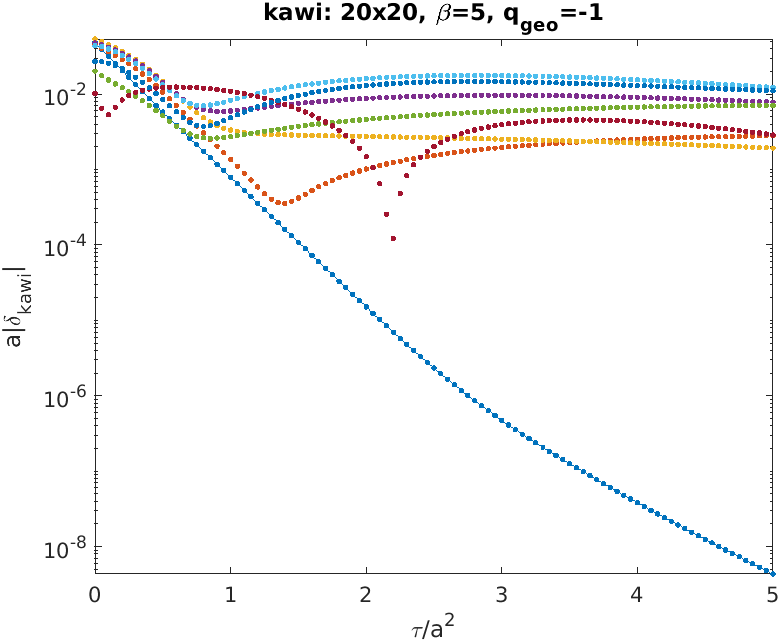}%
\includegraphics[width=0.5\textwidth]{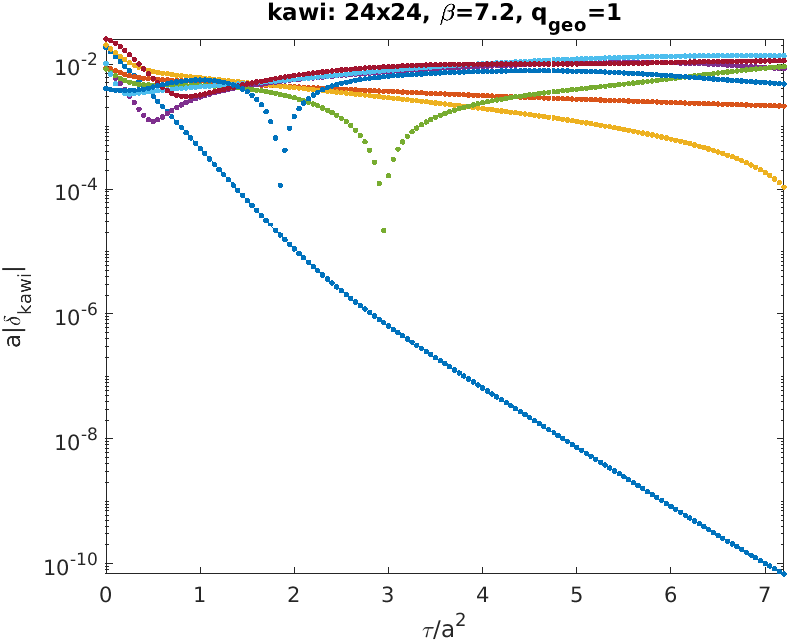}\\
\includegraphics[width=0.5\textwidth]{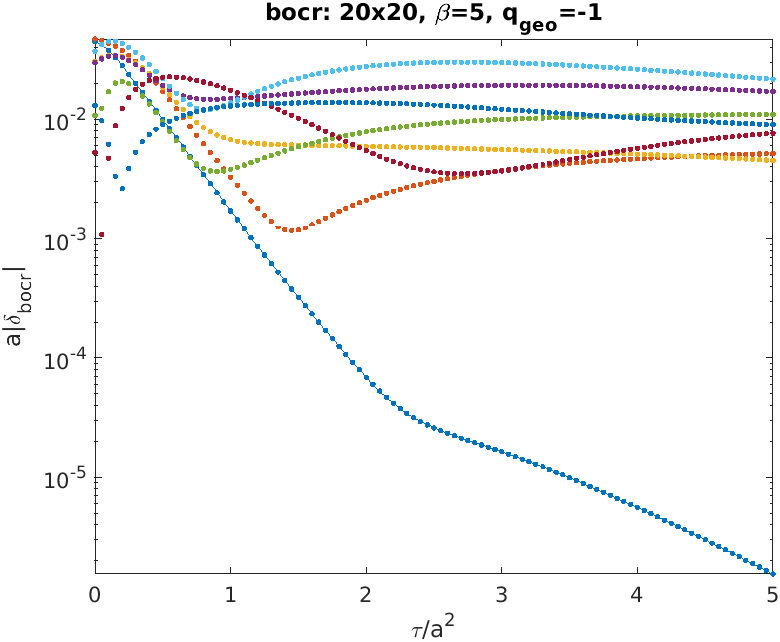}%
\includegraphics[width=0.5\textwidth]{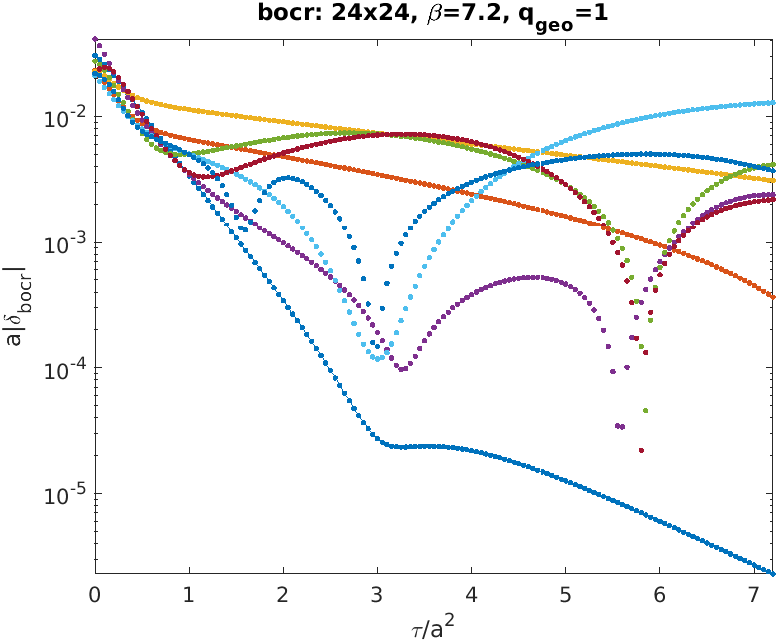}%
\vspace*{-4pt}
\caption{\label{fig:05p0a07p2}
Taste splittings similar to the right-hand panels of Fig.\,\ref{fig:03p2} but for $\be\!=\!5.0$ (left) and $\be\!=\!7.2$ (right).}
\end{figure}

\begin{figure}[!t]
\includegraphics[width=0.5\textwidth]{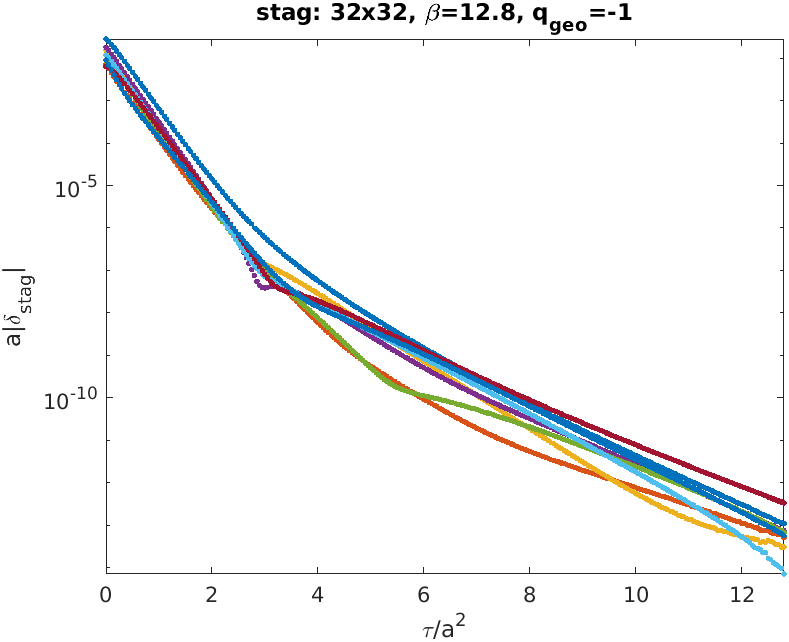}%
\includegraphics[width=0.5\textwidth]{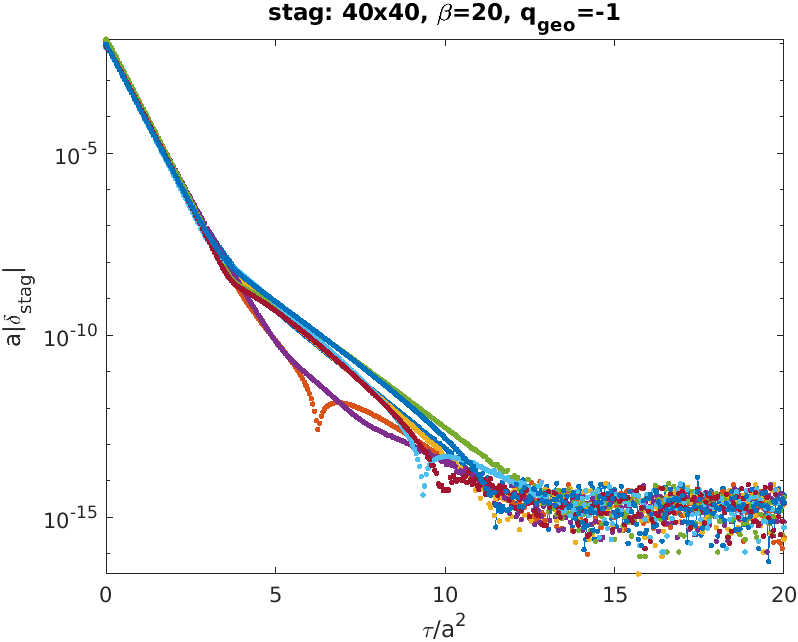}\\
\includegraphics[width=0.5\textwidth]{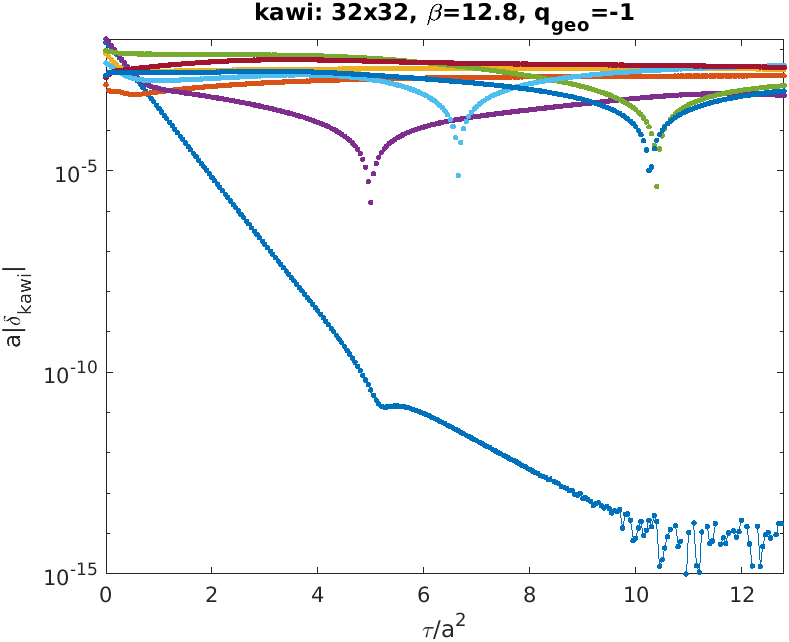}%
\includegraphics[width=0.5\textwidth]{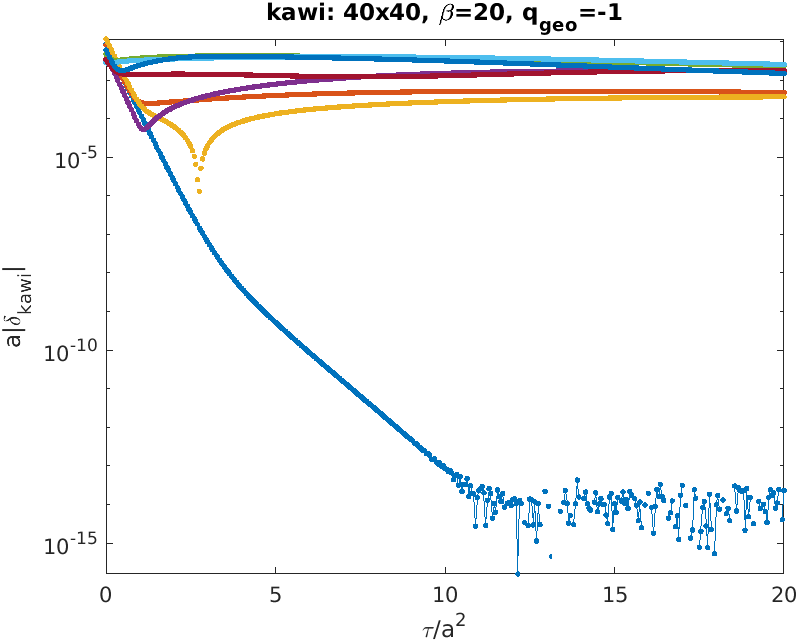}\\
\includegraphics[width=0.5\textwidth]{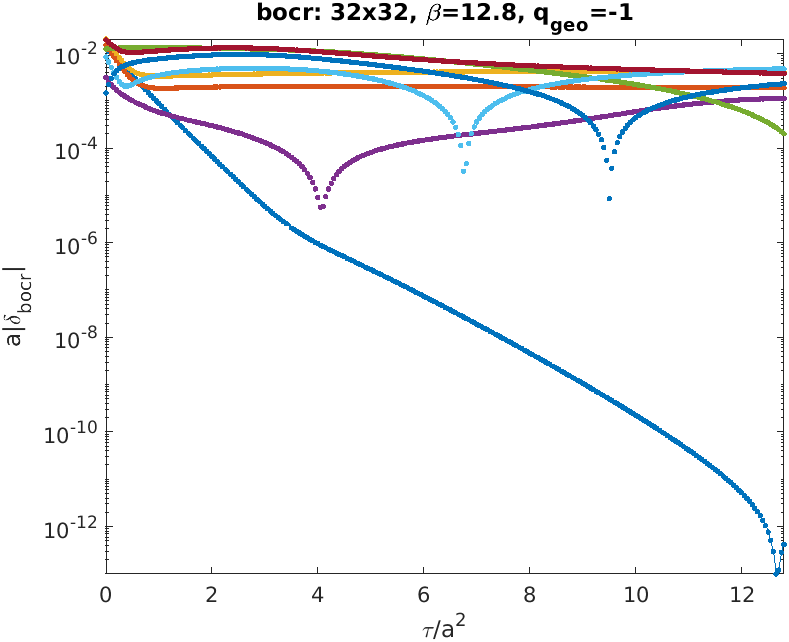}%
\includegraphics[width=0.5\textwidth]{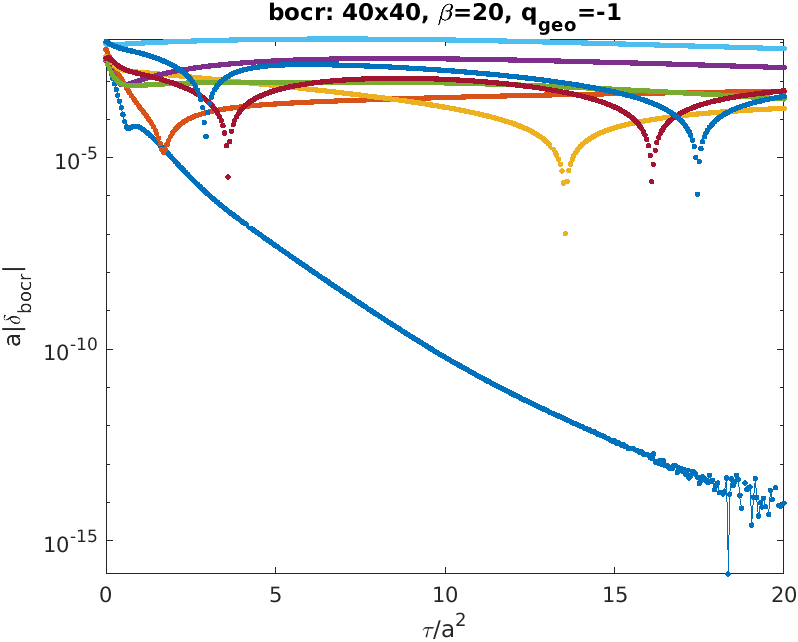}%
\vspace*{-4pt}
\caption{\label{fig:12p8a20p0}
Taste splittings similar to the right-hand panels of Fig.\,\ref{fig:03p2} but for $\be\!=\!12.8$ (left) and $\be\!=\!20.0$ (right).}
\end{figure}

\begin{figure}[!t]
\includegraphics[width=0.5\textwidth]{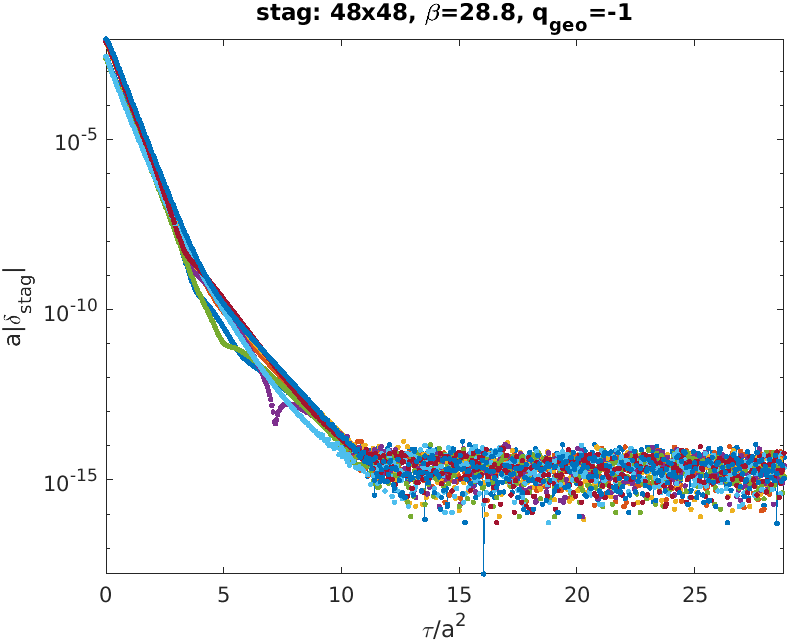}%
\includegraphics[width=0.5\textwidth]{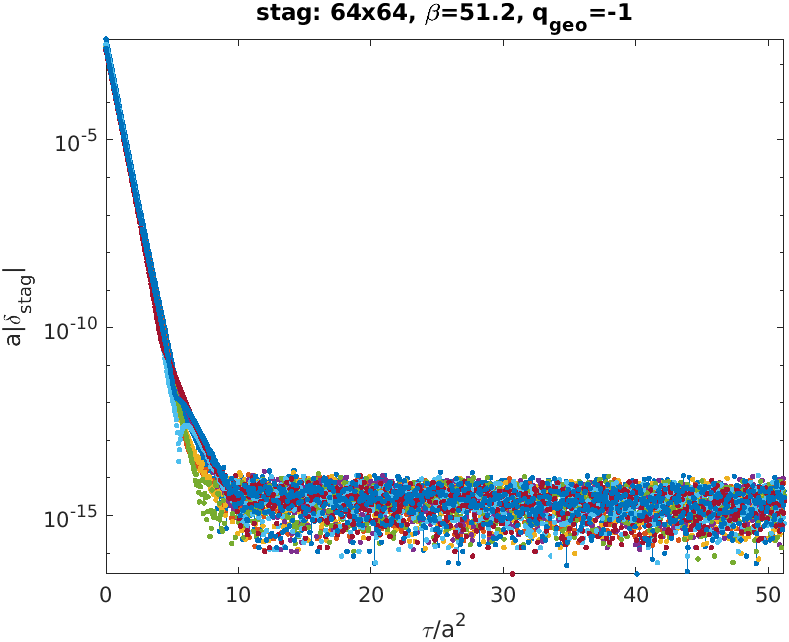}\\
\includegraphics[width=0.5\textwidth]{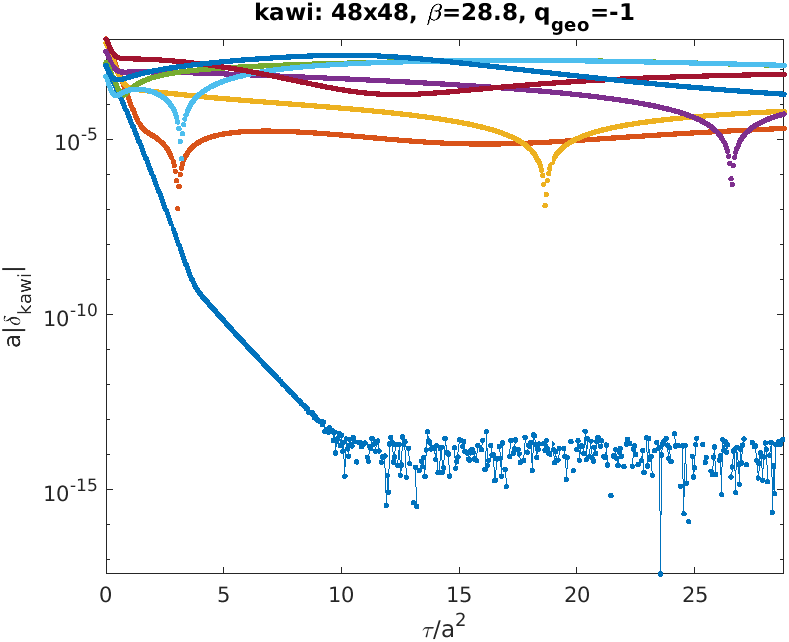}%
\includegraphics[width=0.5\textwidth]{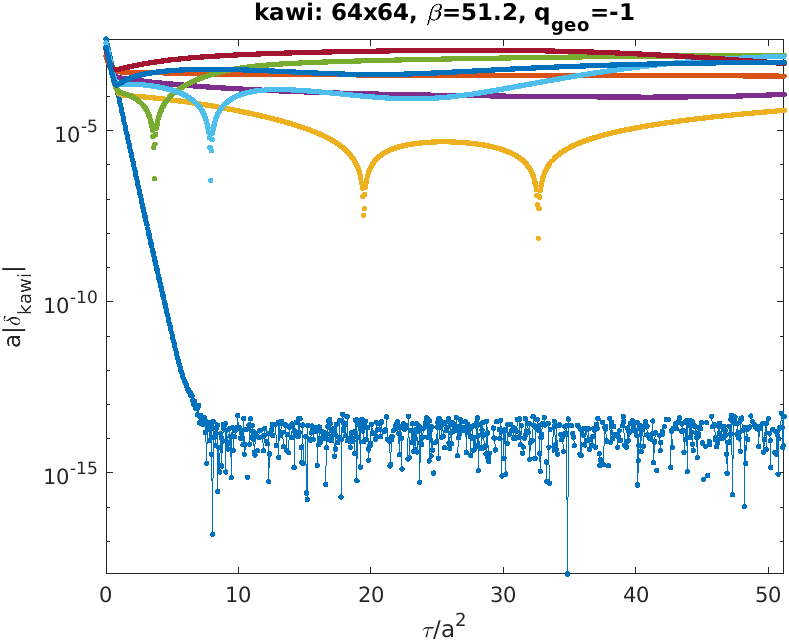}\\
\includegraphics[width=0.5\textwidth]{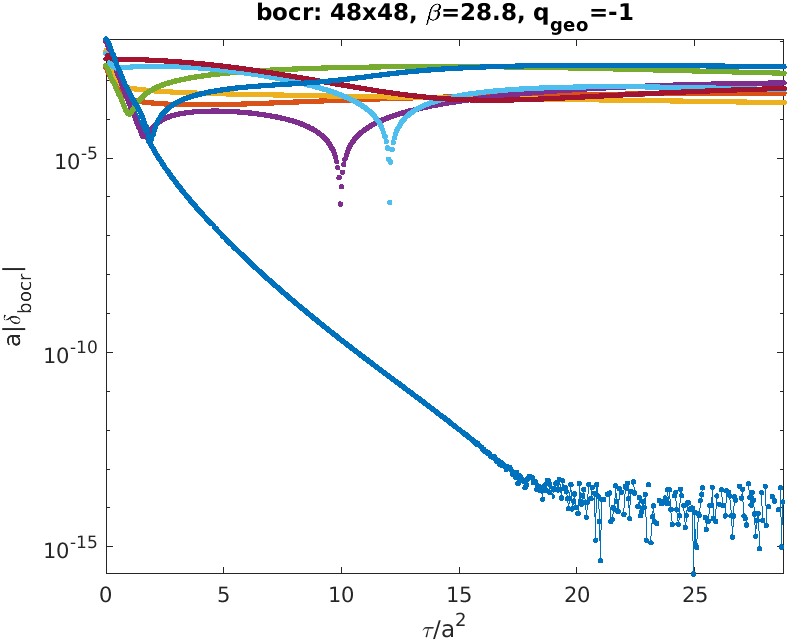}%
\includegraphics[width=0.5\textwidth]{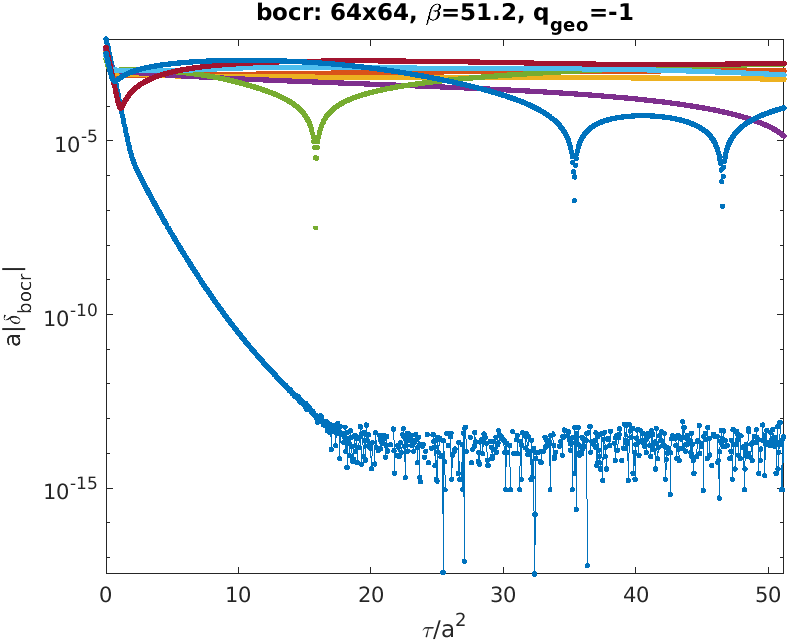}%
\vspace*{-4pt}
\caption{\label{fig:28p8a51p2}
Taste splittings similar to the right-hand panels of Fig.\,\ref{fig:03p2} but for $\be\!=\!28.8$ (left) and $\be\!=\!51.2$ (right).}
\end{figure}

\begin{table}
\centering
\begin{tabular}{|c|ccccccc|}
\hline
$\be$                             &  3.2 &  5.0 &  7.2  &  12.8 &  20.0 &  28.8 &  51.2 \\
\hline
$a|\de_\mr{stag}|$ at $\ta/a^2=1$ & 3e-3 & 1e-3 & 1e-3  & 3e-4  & 3e-4  & 1e-4  & 1e-4  \\
$a|\de_\mr{stag}|$ at $e^2 \ta=1$ & 1e-5 & 1e-8 & 1e-10 & 1e-12 & $\ep$ & $\ep$ & $\ep$ \\
\hline
\end{tabular}
\vspace*{-2pt}
\caption{\label{tab:zeros} Typical size of $a|\de_\mr{stag}|$ at $\ta/a^2=1$ and $e^2 \ta=1$. Here $\ep$ denotes a zero to machine precision.}
\vspace*{-8pt}
\end{table}

\section{Effect of the gradient flow on Dirac operator eigenvalues}

The massless staggered Dirac operator $aD_\mr{stag}$ has purely imaginary eigenvalues which come in pairs $\pm\ri\la$, due to $\ep$-hermiticity.
In Fig.\,\ref{fig:03p2} we plot the 15 smallest imaginary parts $\la>0$ on the original background $U$ at $\ta/a^2=0$.
At this stage, no pairing is visible.
Only as we repeat this for smoothed backgrounds $V$, the pairing becomes visible at $\ta/a^2\simeq1$.
The staggered taste splittings (e.g.\ $\de_1=2\la_1$, $\de_2=\la_3-\la_2$ for $q=1$, see Ref.~\cite{Ammer:2024yro}) all seem to decline exponentially in $\ta/a^2$.
For KW and BC fermions, the situation is similar for the pair $\pm\ri\la_1$ (which is the would-be zero-mode pair for $|q|=1$),
while all non-topological mode splittings diminish only reluctantly.

In Fig.\,\ref{fig:05p0a07p2} we repeat this for $\be=5.0,7.2$, in Fig.\,\ref{fig:12p8a20p0} for $\be=12.8,20.0$, and in Fig.\,\ref{fig:28p8a51p2} for $\be=28.8,51.2$.
Throughout, we select a single representative configuration with topological charge $q=\pm1$.
What changes is the maximum flow-time in lattice units, in line with Tab.\,\ref{tab:planning}.
Beginning at $\be=12.8$ double precision may be insufficient to resolve the smallest taste splitting.

To summarize one may say that increasing $\be$ in a fixed physical volume did not bring any significant change.
The would-be zero-mode splitting decreases roughly exponentially for each formulation.
But the non-topological zero-mode splittings diminish in this way only in the staggered case,
while they reach values $a|\delta_\mr{KW,BC}|\!\simeq\!10^{-3}$ in the KW/BC cases.
For staggered fermions it is interesting to compare $a|\de_\mr{stag}|$ at fixed flow-time in lattice/physical units across $\be$, see Tab.\,\ref{tab:zeros}.


\section{Conclusions}

\vspace*{-4pt}

KW and BC fermions distinguish between would-be zero-mode splittings (which decrease exponentially in the gradient flow time) and non-topological mode splittings (which do not).
By contrast, staggered fermions reduce their taste breakings \emph{exponentially} with the flow time, regardless of the nature of the underlying continuum mode.
The good news for staggered practitioners is that \emph{all} taste splittings disappear when \emph{at least one} of the limits $\be\to\infty,\ta/a^2\to\infty$ is taken.

It will be interesting to extend this investigation to ensembles at various $(\be,L/a)$ combinations, keping the physical box size fixed as in Ref.~\cite{Ammer:2024yro}.
The gradient flow allows for two smoothing strategies: the flow time may be kept fixed in lattice units ($\ta/a^2=\mr{const}$) or
in physical units ($e^2\ta=\mr{const}$ in 2D), see Tabs.\,\ref{tab:planning} and \ref{tab:zeros}.
In the first case, locality of the fermion formulation in the continuum limit is guaranteed by construction (as is true with any fixed number of stout steps).
In the second case, locality is an issue but the lattice regulator gets replaced by a diffusive regulator with
universal properties \cite{Luscher:2010iy,Luscher:2011bx} (see also the discussion in Ref.~\cite{Ammer:2024hqr}).

An issue not addressed so far is the admixture of lower-dimensional operators to $D_\mr{KW}$ and $D_\mr{BC}$
\cite{Capitani:2010nn,Weber:2023kth,Godzieba:2024uki}, as the respective coefficients are not known in 2D.
We plan to embark on such a calculation; with such numbers in hand one can try to compensate these mixing effects.
Perhaps, with correct unmixing, a future version of our KW/BC taste splitting plots might look different\,?

\vspace*{-4pt}



\end{document}